\pdfoutput=1 
\documentclass[useAMS,usenatbib,usegraphicx]{mn2e}

\def\lesssim{\mathrel{\hbox{\rlap{\hbox{\lower4pt\hbox{$\sim$}}}\hbox{$<$}}}}
\def\gtrsim{\mathrel{\hbox{\rlap{\hbox{\lower4pt\hbox{$\sim$}}}\hbox{$>$}}}}

\newcommand{\mamo}[1]{\mbox{$#1$}}
\newcommand{\unit}[1]{\ifmmode \:\mbox{\rm #1}\else \mbox{#1}\fi}
\newcommand{\sbr}[1]{_{\rm #1}}
\newcommand{\mone}{\mamo{^{-1}}}

\newcommand{\mpc}{\unit{Mpc}}

\newcommand{\hmpc}{\mamo{h\mone}\mpc}

\newcommand{\secref}[1]{Section~\ref{sec:#1}}

\newcommand{\Eqref}[1]{Equation~(\ref{eq:#1})}
\newcommand{\figref}[1]{Fig.~\ref{fig:#1}}
\newcommand{\tabref}[1]{Table~\ref{tab:#1}}

\title[Finding galaxy groups with photometric redshifts]{Group-finding with photometric redshifts: The Photo-z Probability Peaks algorithm}
\author[Bryan R. Gillis {\it et al.}]{Bryan R. Gillis$^{1}$\thanks{E-mail:
bgillis@sciborg.uwaterloo.ca}, Michael. J. Hudson$^{1,2,3}$ \\
$^{1}$Department of Physics and Astronomy, University of Waterloo, Waterloo, ON N2L 3G1, Canada.\\
$^{2}$Institut d'Astrophysique de Paris - UMR 7095, CNRS/Universit\'{e} Pierre et Marie Curie, 98bis boulevard Arago, 75014 Paris, France.\\
$^{3}$Perimeter Institute for Theoretical Physics, 31 Caroline St. N., Waterloo, ON, N2L 2Y5, Canada.}
\begin{document}

\date{Jun 8 2010}

\pagerange{\pageref{firstpage}--\pageref{lastpage}} \pubyear{2010}

\maketitle

\label{firstpage}

\begin{abstract}
We present a galaxy group-finding algorithm, the Photo-z Probability Peaks (P3) algorithm, optimized for locating small galaxy groups using photometric redshift data by searching for peaks in the signal-to-noise of the local overdensity of galaxies in a three-dimensional grid. This method is an improvement over similar two-dimensional matched-filter methods in reducing background contamination through the use of redshift information, allowing it to accurately detect groups at lower richness. We present the results of tests of our algorithm on galaxy catalogues from the Millennium Simulation. Using a minimum S/N of 3 for detected groups, a group aperture size of 0.25 $h^{-1}$Mpc, and assuming photometric redshift accuracy of $\sigma_z = 0.05$ it attains a purity of $84\%$ and detects $\sim 295$ groups/deg.$^2$ with an average group richness of 8.6 members. Assuming photometric redshift accuracy of $\sigma_z = 0.02$, it attains a purity of $97\%$ and detects $\sim 143$ groups/deg.$^2$ with an average group richness of 12.5 members. We also test our algorithm on data available for the COSMOS field and the presently-available fields from the CFHTLS-Wide survey, presenting preliminary results of this analysis.
\end{abstract}


\section{Introduction}

Most galaxies in the Universe are gravitationally bound to one or more other galaxies within galaxy groups. A number of recent studies have indicated that the mass-to-light ratios of groups may be a steep function of the group mass\citep{MarHud02, EkeFreBau04, ParHudCar05, WeivanYan06}. This phenomenon may be due to the presence of a critical halo mass above which star formation is efficiently quenched \citep{DekBir06, GilBal08}. Clearly, it is of interest to improve existing data on the mass-to-light ratio on the mass scale close to that of groups, in order to better determine whether, for example, such a critical halo mass exists, and, more generally, to determine what mechanisms may be responsible for the quenching of star formation in the group environment.

The analysis of the mass-to-light ratios of poor groups has been limited by the difficulty both in identifying them, and in estimating their masses. The identification of groups is typically based on spectroscopic redshifts, with galaxies assigned to groups through methods such as the Friends-of-Friends algorithm \citep{HucGel82}. There are several methods to estimate the masses of groups: X-ray-derived masses, a method that is limited to rich groups \citep{MulZab98}; virial estimates based on redshifts; and weak gravitational lensing (WL).

Virial mass estimators are problematic for small groups: the virial mass estimator scales as $\sigma^2$, where $\sigma$ is the velocity dispersion of the group. The accuracy of this method is tied tightly to the number of galaxies in the group; for example,  for $N\sbr{mem}=6$, the estimated $\sigma$ is uncertain to a factor of $\sim2$, leading to large uncertainty in the virial mass \citep{KnoLilIov09}. More importantly, the estimator assumes that the group has reached dynamic equilibrium and that the orbital velocity anisotropy is known; if these assumptions are incorrect, it may lead to a systematic bias.

WL has an advantage over virial estimators because the mass estimates are independent of the current dynamical state of the group. WL mass estimates are particularly valuable for poor groups, for which X-ray-derived masses are unobtainable, and virial estimates are most uncertain. However, the signal-to-noise ($S/N$) for a single group is very low,  so it is necessary to ``stack'' the signal from many groups. Furthermore, the lensing mass is sensitive to all overdensities along the line of sight, so this requires careful calibration with simulations. Previous weak lensing studies of poor systems include \cite{HoeFraKui01, ParHudCar05, SheJohMas09} who studied samples of 59, 116, and 132 473 systems respectively.

This paper is the first in a series based on data from the CFHTLS-Wide survey \citep{CFHTLSwide}, presenting the method we will be using to identify groups in the CFHTLS-Wide. In future work, we expect to use weak lensing to estimate the masses of groups in the CFHTLS-Wide. To date there has been no spectroscopic survey of the entire 170 square degrees of the CFHTLS-Wide, so here we will use photometric redshifts to assign galaxies to groups.  Photometric redshifts have significantly larger random errors than spectroscopic redshifts ($> 0.02$ versus $\lesssim 0.001$) \citep{Ben00}. Due to the large photometric redshift errors, any identified groups will suffer significant contamination from field galaxies. These projection effects will need to be carefully calibrated and corrected when estimating group richness. 

Previous methods of detecting groups and clusters using only photometric data have focused on clusters, and/or on red-sequence galaxies. The cluster-red-sequence method \citep{GlaYee00,LuGilBal09}, which searches for overdensities of red-sequence galaxies, is optimized for clusters with more than 20 red-sequence galaxies, significantly above the mass scale of interest for this project. The K2 method developed by \citet{ThaWilCra09} can generate a catalog of $\sim 99\%$ purity, with reasonable completeness for poor clusters. However, the method assumes that galaxies in the same group will have very similar colours. This assumption may result in a significant bias toward groups where the galaxies are all at a similar stage of evolution, particularly among poor groups. The probability-friends-of-friends algorithm \citep{LiYee08} is better suited to the requirements of this project, producing $\sim 90\%$ purity for groups with 8 or more members, but the purity decreases rapidly below this point, to $\sim 70\%$ for groups with 5 or more members.

With all group/cluster-finding methods there is a trade-off between purity and completeness \citep{KnoLilIov09}. We expect to have a very large sample of groups, and so can afford to sacrifice completeness for purity in our group selection.  Although the probability-friends-of-friends algorithm might be usable with some refinements, our method has shown to be able to produce significantly higher purity ($\sim 80\%-\sim 95\%$ for groups of 5 or more members, depending on the quality of the photometry), albeit at the expense of completeness. With the large quantity of data available in the CFHTLS-Wide survey, this is an acceptable trade-off.

Our method, which we refer to as the Photo-z Probability Peaks (P3) algorithm, involves finding peaks in three dimensions using the photo-z probability distribution functions (PDFs).  This method is similar in spirit to the cluster-finding algorithms recently published by \citet{MilWaeHey10} and \citet{AdaDurBen10}. Whereas their methods are optimizing for detecting clusters with high completeness, our method is tuned to finding groups, and assembling a group catalogue that has high purity.

In Section 2 of this paper we explain our method for identifying galaxy groups. Section 3 gives the results of tests of our method on simulated and observed data sets, for different choices of the algorithms parameters and for different assumptions regarding the accuracy of the photometric redshifts. Section 4 discusses the applicability of this method to the galaxy catalogues from the CFHTLS-Wide survey, including preliminary results and comparisons with other group catalogues made from these fields.  Throughout we adopt a cosmology with the following parameters: $\Omega_m = 0.3$, $\Omega_{\Lambda} = 0.7$, and $H_0 = 70 \mathrm{ km/s/Mpc}$. All magnitudes are in the AB system unless stated otherwise.

\section{Group Finding Method}

The methodology behind the P3 algorithm involves searching for significant overdensities in the distribution of galaxies in 3Ds. Specifically, to search for overdensities, we construct a three-dimensional grid of points within the lightcone of field to be analyzed, and at each point, we calculate the local overdensity of galaxies in a circular aperture surrounding the point, and compare this to the nearby background in an annulus surrounding the point.

In practice, we adopt a 3D grid with a spacing of $R\sbr{g}=\sim 0.2$ comoving $h^{-1}$Mpc in the transverse direction, with each redshift slice having a thickness of $z\sbr{g} =  0.02$. A small galaxy group will have a radius of $\sim 0.25$ $h^{-1}$Mpc, so it is resolved with this spacing. The typical photo-z errors are $\sim 0.04$, so are also resolved with this spacing. However, high quality photometric redshifts, such as those provided by \citet{IlbCapSal09} may have lower errors and require a finer grid-spacing. 

Our calculation for the galaxy surface density within the aperture (represented by $\rho\sbr{ap}$, though note that the calculated density is only pseudo-3D, as we use a probability density in the $z$-dimension). The method is designed to handle regions of the sky which have been masked (e.g. due to bright stars). 
The  procedure is as follows:
\begin{itemize}
\item For each galaxy, use the photometric redshift probability density function -- here determined by via the Bayesian Photometric Redshift Estimation (BPZ) code \citep{Ben99} ) -- to obtain the probability ($p_i$) that it is within a given redshift slice of thickness $z\sbr{g}$.
\begin{itemize}
\item  In this paper, our algorithm approximates the PDF as a Gaussian distribution to decrease computation time (but can easily be generalized to arbitrary PDFs).
\item We multiply the redshift probability calculated above by the BPZ ODDS parameter provided by the photometric redshift method. In BPZ, the ODDS parameter gives the probability that the true redshift lies within the primary peak of the PDF, and so is necessary in normalizing the Gaussian height.
\end{itemize}
\item Sum the weighted probabilities for all galaxies within the aperture and divide by the area of the aperture which falls outside any masked regions ($A\sbr{ap}$). This gives us the density within the aperture, as in \Eqref{rhoap}.
\end{itemize}
\begin{equation}
\label{eq:rhoap}
\rho\sbr{ap} = \frac{\Sigma_{i=1}^n w_i p_i}{A\sbr{ap}}
\end{equation}
This procedure can also be used with only minor modifications to determine the density within the annulus surrounding the test point, which will give us the local background galaxy density, and thus the overdensity $\delta$. For our purposes, we use an annulus with inner radius of $1\,h^{-1}$Mpc and outer radius $3\,h^{-1}$Mpc. This large size minimizes the effect of large-scale structure such as superclusters, filaments, and walls on the calculated $\delta$, while it remains small enough to be representative of density variation caused by observational effects.

In order to obtain a pure sample of groups, we will select only groups with a sufficiently high signal-to-noise ratio (S/N) in $\delta$. In order to calculate the noise in our measurement of $\delta$, we model the number of galaxies which make a significant contribution to the density as a Poisson distribution. In the procedure above, we include only galaxies that have a probability of being within this redshift slice of at least $0.1\%$, rounding other probabilities down to zero. Using Poisson statistics, a sample which finds $n$ contributing galaxies would give us a standard error of $\sqrt{n}$. We can then estimate the Poisson error of our density as:
\begin{equation}
\sigma\sbr{ap,Poisson} = \frac{\left< w_i p_i \right> * \sqrt{n}}{A\sbr{ap}} = \frac{\rho\sbr{ap}}{\sqrt{n\sbr{ap}}}
\end{equation}
We can then repeat these calculations for the annulus, giving the error in its density. In the end, we combine these errors in quadrature to give the final noise in $\delta$. This allows us to calculate the S/N for each test point.

With our 3D grid of S/N, we then proceed to detect the peaks, as these are most likely to correspond to the centres of galaxy groups. In order to not identify multiple peaks with the same group, we apply a threshold distance of $R\sbr{t}=0.5\,h^{-1}$Mpc, the size of a large group, and a threshold of $z\sbr{t}=0.02$ in redshift in which a peak must be the highest point, rather than simply requiring that the peak must be higher than the points immediately surrounding it in the grid. This procedure minimizes the chance of groups being detected at multiple points in the sky, as any substructure that lies within $R\sbr{t}$ of the group centre will be ignored. The S/N of a rich group also tends to steadily increase toward the centre, so even substructure more distant than $R\sbr{t}$ from the centre is likely to lie within $R\sbr{t}$ of at least one point with greater S/N, and thus it will also be ignored. Multiple detections along the line of sight are more difficult to handle, as the scatter of photometric redshifts within a rich cluster can approach 0.2 \citep{AdaIlbPel08}. However, this is primarily a problem with richer groups, as a large number of galaxies is necessary for multiple peaks to be observed at redshifts separated by more than $z\sbr{t}=0.02$. Future refinements to the algorithm may work to address this issue by merging multiple peaks that lie along the line of sight.

Once the peak catalogue is complete, we then extract only those peaks above some signal-to-noise threshold. This leaves us with our ultimate group catalogue.  The detail of how the signal-to-noise thresholds and the aperture sizes are chosen are described in the following section.

\section{Tests of the Group Finding Algorithm}

In order to test the P3 algorithm, we compared its results to a catalogue of dark matter halos containing at least two galaxies in six lightcones extracted from the Millennium simulation by \citet{KitWhi07}, and to a friends-of-friends spectroscopic group catalogue generated by \citet{KnoLilIov09} using the zCOSMOS 10k sample covering the COSMOS field, which overlaps with the CFHTLS D2 field. 

\subsection{Comparison with Simulations}

To assess the accuracy of the P3 algorithm against an ideal catalogue, we used six simulated 2 deg.$^2$ lightcones extracted from the Millennium simulation \citep{SprWhiJen05,DeLBla07} by \citet{KitWhi07}. Given the resolution limits of the Millennium simulation, the catalogue is complete for Johnson {\it I} $<$ 24 in the AB system. We also used a magnitude limit of {\it I} $<$ 22.5 in the {\it I}-band for most of the testing, as this matches the spectroscopic catalogue of \citeauthor{KnoLilIov09} (see \secref{realdat} below). We also tested including galaxies with {\it I}-band magnitudes between 22.5 and 24 to assess how this affected our accuracy (see \secref{faintdat}, also below). For our testing, we used only galaxies between $z\sbr{lo}=0.2$ and $z\sbr{hi}=0.8$, as this is where we expect to attain the best lensing signal.

To simulate photometric redshifts for this galaxy catalogue, for simplicity we applied a Gaussian deviate to the redshifts of the galaxies. We generated two mock photo-z catalogues, each using different simulated photo-z errors. The first mock catalogue, hereafter CFHTLSpz, simulated the accuracy of the photometric redshifts in the CFHTLS Deep fields of \citet{IlbArnMcC06}, with a redshift error of 0.05 for {\it I} $<$ 22.5, and 0.10 for 22.5 $<$ {\it I} $<$ 24. The second catalogue, hereafter COSMOS30pz, mimicked the accuracy of the COSMOS-30 \citep{IlbCapSal09} photometric redshifts: 0.02 for {\it I} $<$ 22.5, and 0.04 for 22.5 $<$ {\it I} $<$ 24.  We note that, after these tests were done, a recent analysis of photo-z's in the CFHTLS-Wide survey \citep{HilPieErb09} suggests that the errors in this survey will be approximately 0.03 for {\it I} $<$ 22.5 and $0.2 < z < 1.1$, which lies between the two error ranges tested in this paper.

Catastrophic errors, where the actual redshift of the galaxy differs from its photometric redshift by many standard deviations, were not simulated in our tests. With real data, we will be able to select only galaxies that have a minimal chance of being catastrophic errors. This selection will likely result in a slightly less complete catalogue, but should guarantee that the purity is not decreased due to catastrophic errors. Within the redshift range of $0.2 < z < 0.8$, we expect the fraction of catastrophic errors to be less than $5\%$ \citep[][Hildebrandt {\it et al.} in preparation]{CouIlbKil09}.

The Millennium simulation also contains halo information for galaxies. We used halos identified by \citet{DeLSprWhi06} through a friends-of-friends method applied to the dark matter particles to identify the real groups of galaxies. The halo centres were determined to be the average positions of the galaxies contained in the halo. Only halos containing at least two galaxies were used in the group comparisons. In total, we obtained 31668 groups across the six fields, for approximately 2600 groups per deg.$^2$

\begin{figure*}
  \centering
  \includegraphics[scale=0.5]{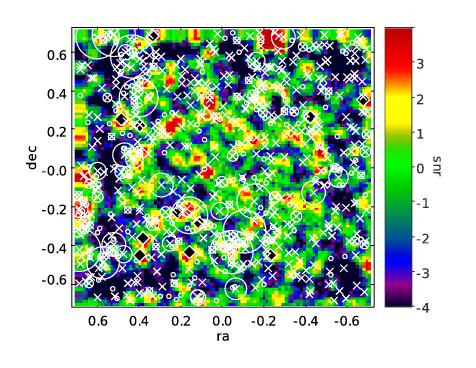}
  \includegraphics[scale=0.5]{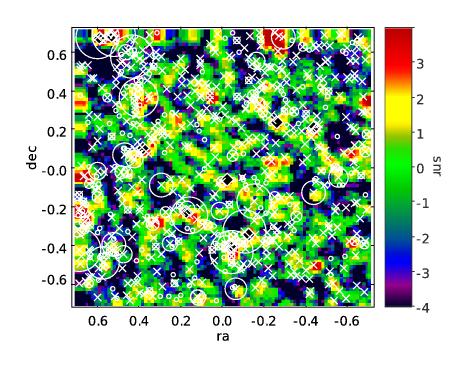}
  \includegraphics[scale=0.5]{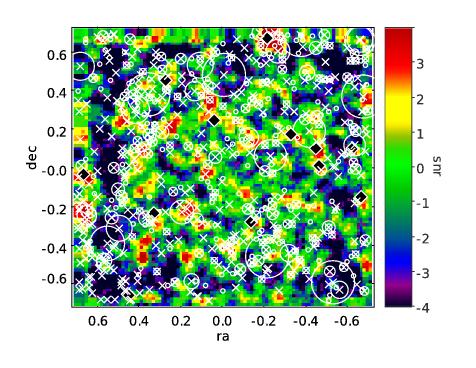}
  \includegraphics[scale=0.5]{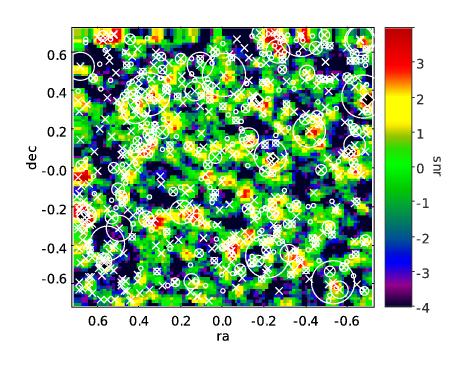}
  \includegraphics[scale=0.5]{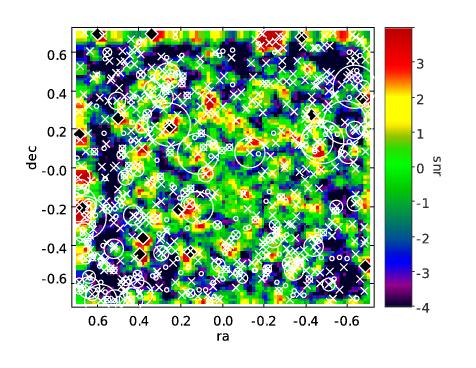}
  \includegraphics[scale=0.5]{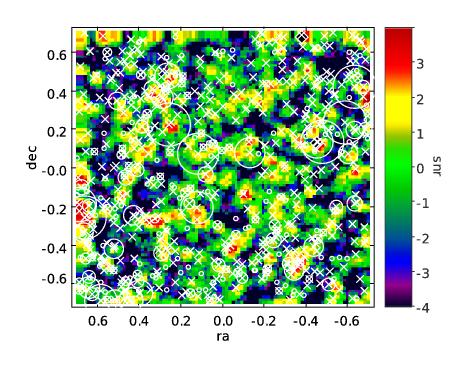}
  \caption[S/N plots in R.A. and Dec. for data from the Millennium simulation.]{The calculated S/N for the $\delta$ of galaxies on a grid of points in R.A. and Dec., sliced at different values of the redshift, for a field drawn from the Millennium simulation by \citet{KitWhi07}. S/N is indicated by the colour. Locations of halos with at least 3 galaxies are indicated by white circles, with their radii indicating the richnesses of the groups. White crosses indicate the location of a circle in a nearby layer, within $\Delta z = 0.04$. Detected peaks with a S/N $>$ 3 are indicated by the black diamonds. Peaks are detected in three dimensions, so what appear to be peaks in the individual 2D plots may actually be detected on another slice. Additionally, peaks have a threshold radius, $R\sbr{t}=0.5\,h^{-1}$Mpc, within which they must be the highest point to count as a peak, so some peaks may not be detected if they are sufficiently close to another peak. Left column contains plots using CFHTLSpz errors with $R\sbr{ap} = 0.5\,h^{-1}$Mpc, right column contains plots using COSMOS30pz errors with $R\sbr{ap} = 0.5\,h^{-1}$Mpc. Redshift slices, from top to bottom: 0.58, 0.60, 0.62.}
  \label{fig:millsnrz}
\end{figure*}
\begin{figure*}
  \centering
  \includegraphics[scale=0.5]{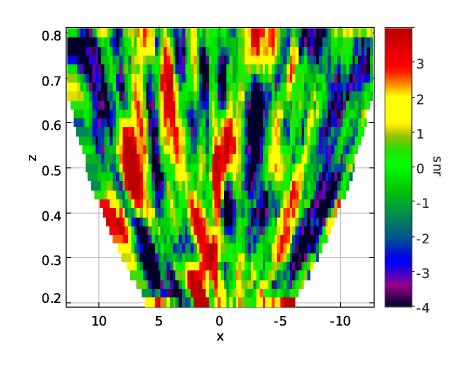}
  \centering
  \includegraphics[scale=0.5]{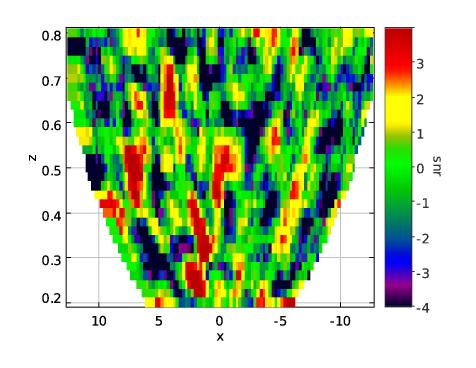}
  \centering
  \includegraphics[scale=0.5]{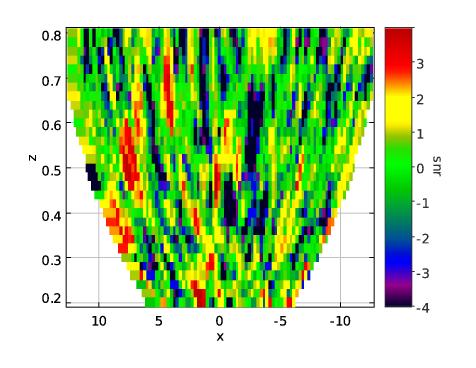}
  \centering
  \includegraphics[scale=0.5]{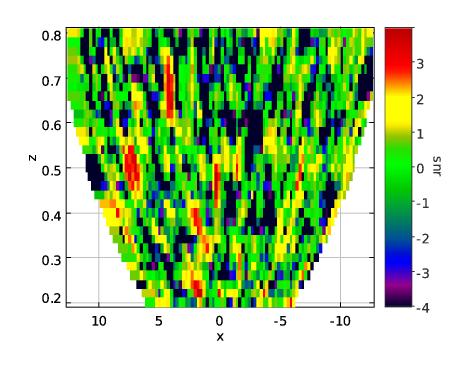}
  \centering
  \caption[S/N plots in x and z for data from the Millennium simulation.]{An alternate view of the plots from \figref{millsnrz}, sliced at constant Dec., showing how far groups extend in the redshift dimension. The left column shows simulations with CFHTLSpz errors, and the right column shows simulations with COSMOS30pz errors. The top row shows simulations with $R\sbr{ap} = 0.5\,h^{-1}$Mpc, and the bottom row shows simulations with $R\sbr{ap} = 0.25\,h^{-1}$Mpc.}
  \label{fig:millsnry}
\end{figure*}

\figref{millsnrz} and \figref{millsnry} show a graphical representation of the S/N calculated by the P3 algorithm for a selection of redshift slices, to illustrate how the detected peaks correspond to actual groups. Recall that peaks are detected in three dimensions, so what appear to be peaks in the individual 2D plots may actually be detected on another slice. Additionally, peaks have a threshold radius $R\sbr{t}=0.5\,h^{-1}$ Mpc within which they must be the highest point to count as a peak, so some peaks may not be detected if they are sufficiently close to another peak.

\subsubsection{Purity and Completeness}

\begin{table*}
 \centering
  \caption[Summary of group-finding accuracy for data from the Millennium simulation.]{Summary of the purity and completeness of the P3 algorithm when its results are matched to the Millenium halo catalogue, containing 31668 total groups, for various minimum S/N limits and a Control. The combined galaxy catalogues cover a total of 12 deg.$^2$ of simulated sky. Cut is what, if any, S/N limit is applied to the peak catalogue; $R\sbr{ap}$ is the aperture radius in $h^{-1}$Mpc; Nhit is the number of detected peaks which match to a halo; Npeak is the total number of detected peaks; P is the fraction of peaks which match to a halo; C is the fraction of halos which match to a detected peak; $\left<N\sbr{m}\right>$ is the mean number of members in the detected peaks which match to a halo; $\left<m\right>$ is the geometric mean mass of the detected peaks which match to a halo, in units of $10^{12}\,h^{-1}M_{\odot}$; $\left<N\sbr{mat}\right>$ is the mean number of halos within the matching distance of a peak (not including peaks which match to zero halos); and $\left<N\sbr{field}\right>$ is the mean number of field galaxies within the matching distance of a peak (also not including peaks which match to zero halos).}
  \resizebox{\textwidth}{!}{
  \begin{tabular}{ l r | r r r r r r r r | r r r r r r r r }
  \hline
   & & \multicolumn{8}{c|}{CFHTLSpz errors} & \multicolumn{8}{c}{COSMOS30pz errors} \\
  \hline
  Cut & $R\sbr{ap}$ & Nhit & Npeak & P & C & $\left<N\sbr{m}\right>$ & $\left<m\right>$ & $\left<N\sbr{mat}\right>$ & $\left<N\sbr{field}\right>$ & Nhit & Npeak & P & C & $\left<N\sbr{m}\right>$ & $\left<m\right>$ & $\left<N\sbr{mat}\right>$ & $\left<N\sbr{field}\right>$ \\
  \hline
Control & 0.5 & 1705 & 2988 & 0.57 & N/A & 3.87 & 1.37 & 1.82 & 6.64 & 1705 & 2988 & 0.57 & N/A & 3.87 & 1.37 & 1.82 & 6.64 \\
All Peaks & 0.5 & 2502 & 3439 & 0.73 & 0.18 & 5.64 & 2.35 & 2.13 & 6.84 & 2493 & 2805 & 0.89 & 0.21 & 6.81 & 2.84 & 2.41 & 7.65 \\
S/N $>$ 2 & 0.5 & 1839 & 2406 & 0.76 & 0.13 & 6.27 & 2.42 & 2.23 & 7.20 & 1935 & 2048 & 0.94 & 0.17 & 7.87 & 3.19 & 2.61 & 8.17 \\
S/N $>$ 3 & 0.5 & 1196 & 1496 & 0.80 & 0.09 & 7.20 & 2.41 & 2.38 & 7.61 & 1011 & 1025 & 0.99 & 0.10 & 11.56 & 3.83 & 3.06 & 9.40 \\
S/N $>$ 4 & 0.5 & 607 & 725 & 0.84 & 0.05 & 9.29 & 2.34 & 2.63 & 8.38 & 411 & 415 & 0.99 & 0.05 & 19.65 & 5.09 & 3.52 & 10.52 \\
\hline
Control & 0.25 & 1705 & 2988 & 0.57 & N/A & 3.87 & 1.37 & 1.82 & 6.64 & 1705 & 2988 & 0.57 & N/A & 3.87 & 1.37 & 1.82 & 6.64 \\
All Peaks & 0.25 & 12341 & 20727 & 0.60 & 0.65 & 4.49 & 1.92 & 1.76 & 5.85 & 12119 & 18724 & 0.65 & 0.68 & 4.54 & 1.95 & 1.80 & 6.00 \\
S/N $>$ 2 & 0.25 & 7937 & 10904 & 0.73 & 0.44 & 5.43 & 2.12 & 1.96 & 6.52 & 6426 & 7160 & 0.90 & 0.39 & 6.21 & 2.51 & 2.16 & 7.16 \\
S/N $>$ 3 & 0.25 & 2970 & 3537 & 0.84 & 0.18 & 8.55 & 2.40 & 2.34 & 7.81 & 1670 & 1715 & 0.97 & 0.12 & 12.51 & 3.76 & 2.85 & 9.09 \\
S/N $>$ 4 & 0.25 & 869 & 986 & 0.88 & 0.06 & 15.29 & 2.75 & 2.74 & 9.29 & 441 & 448 & 0.98 & 0.04 & 25.40 & 5.03 & 3.17 & 10.61 \\
  \hline
  \end{tabular}}
 \label{tab:mill}
\end{table*}

In order to assess purity and completeness, it is first necessary to define what constitutes a match. Our comparison method aimed primarily to assess the purity of our samples, so a peak is defined as a match if it lies within a redshift difference of $z\sbr{mat} = 0.04$ and a projected radius $R\sbr{mat} = 0.5\,h^{-1}$Mpc of at least one group-containing halo. These parameters were adopted because $z\sbr{mat}$ is approximately twice the uncertainty in the mean photometric redshift for a group of 5 members, and $R\sbr{mat}$ is approximately the upper size limit for a group. The comparisons were made for peaks selected above various S/N, as well as a ``Control'' sample which consisted of positions generated from a uniform random grid in R.A., Dec., and $z$. The purity, $P$ is then defined as the fraction of P3-peaks that match to a Millennium group-halo.

Although completeness is not the primary goal of P3, we nevertheless measured completeness, $C$, by calculating the number of spectroscopically-identified groups in the field which had at least one P3 group matched to it. \tabref{mill} shows a summary of the accuracy of the P3 algorithm when run on a mock galaxy catalogue from the Millennium simulation, using the catalogue of groups with at least 2 members as a comparison. For a fiducial minimum $S/N$ limit of 3 and $R\sbr{ap}=0.25 \hmpc$, the P3 algorithm typically detects around 295 groups/deg.$^2$ in the redshift range $0.2 < z < 0.8$ for the simulated galaxy catalogue. Of these detected groups, approximately $84\%$ match to at least real group with at least two bright members when we simulate CFHTLSpz errors. This will give us approximately 248 correct group detections per deg.$^2$. Our completeness is very low, however, picking up at best $44\%$ of groups when we use a S/N cut of 2 and $R\sbr{ap} = 0.25\,h^{-1}$ Mpc. For our purposes, this is not a concern, as the total number of groups we detect in the CFHTLS-Wide will be enough for our goals.

When we simulate COSMOS30pz errors, also using a minimum $S/N$ limit of 3 and $R\sbr{ap}=0.25 \hmpc$, the purity increases to $97\%$, with $143$ groups/deg.$^2$ detected and $139$ of these being real. 

\subsubsection{Influence of resolution and matching radius}

\begin{table*}
 \centering
  \caption[Summary of group-finding accuracy for halo catalogue from the Millennium simulation, using a higher resolution and a lower matching length.]{Summary of the purity and completeness of the P3 algorithm when its results are matched to the Millennium halo catalogue, containing 31668 total groups, for various minimum S/N limits and a Control. Here the P3 algorithm used a smaller grid spacing, $R\sbr{g}=\sim0.1\,h^{-1}$Mpc and $z\sbr{g}=0.1$, and $R\sbr{mat}$ was reduced to $0.25\,h^{-1}$Mpc. The combined galaxy catalogues cover a total of 12 deg.$^2$ of simulated sky. Columns are as \tabref{mill}.}
  \resizebox{\textwidth}{!}{
  \begin{tabular}{ l r | r r r r r r r r | r r r r r r r r }
  \hline
   & & \multicolumn{8}{c|}{CFHTLSpz errors} & \multicolumn{8}{c}{COSMOS30pz errors} \\
  \hline
  Cut & $R\sbr{ap}$ & Nhit & Npeak & P & C & $\left<N\sbr{m}\right>$ & $\left<m\right>$ & $\left<N\sbr{mat}\right>$ & $\left<N\sbr{field}\right>$ & Nhit & Npeak & P & C & $\left<N\sbr{m}\right>$ & $\left<m\right>$ & $\left<N\sbr{mat}\right>$ & $\left<N\sbr{field}\right>$ \\
  \hline
Control & 0.5 & 649 & 2988 & 0.22 & N/A & 3.73 & 1.36 & 1.20 & 1.91 & 649 & 2988 & 0.22 & N/A & 3.73 & 1.36 & 1.20 & 1.91 \\
All Peaks & 0.5 & 1080 & 3213 & 0.34 & 0.04 & 9.64 & 3.09 & 1.25 & 1.84 & 974 & 2415 & 0.40 & 0.04 & 11.89 & 4.12 & 1.29 & 1.89 \\
S/N $>$ 2 & 0.5 & 915 & 2493 & 0.37 & 0.03 & 10.49 & 3.15 & 1.27 & 1.94 & 838 & 1810 & 0.46 & 0.03 & 13.28 & 4.56 & 1.32 & 1.99 \\
S/N $>$ 3 & 0.5 & 651 & 1622 & 0.40 & 0.02 & 12.56 & 3.54 & 1.32 & 2.06 & 588 & 1089 & 0.54 & 0.02 & 16.81 & 5.24 & 1.37 & 2.16 \\
S/N $>$ 4 & 0.5 & 386 & 815 & 0.47 & 0.01 & 15.60 & 3.62 & 1.41 & 2.28 & 320 & 502 & 0.64 & 0.01 & 25.26 & 7.20 & 1.46 & 2.35 \\
\hline
Control & 0.25 & 649 & 2988 & 0.22 & N/A & 3.73 & 1.36 & 1.20 & 1.91 & 649 & 2988 & 0.22 & N/A & 3.73 & 1.36 & 1.20 & 1.91 \\
All Peaks & 0.25 & 6860 & 17263 & 0.40 & 0.24 & 5.44 & 2.42 & 1.32 & 2.24 & 7203 & 13326 & 0.54 & 0.27 & 5.77 & 2.58 & 1.38 & 2.56 \\
S/N $>$ 2 & 0.25 & 5299 & 11300 & 0.47 & 0.18 & 6.18 & 2.58 & 1.36 & 2.39 & 5335 & 7994 & 0.67 & 0.19 & 6.84 & 2.93 & 1.46 & 2.77 \\
S/N $>$ 3 & 0.25 & 2639 & 4798 & 0.55 & 0.09 & 8.39 & 2.73 & 1.45 & 2.69 & 2043 & 2628 & 0.78 & 0.07 & 11.57 & 4.20 & 1.66 & 3.21 \\
S/N $>$ 4 & 0.25 & 976 & 1606 & 0.61 & 0.03 & 13.43 & 3.03 & 1.53 & 2.93 & 620 & 746 & 0.83 & 0.02 & 22.31 & 6.17 & 1.77 & 3.47 \\
  \hline
  \end{tabular}}
 \label{tab:mill225}
\end{table*}

Despite the apparently high purity of our results above, there may be some issues with overmatching. The typical size for a poor galaxy group is $\sim0.25\,h^{-1}$Mpc, so we would not expect a S/N peak that corresponds to this galaxy to be separated from the galaxy's center by more than this amount. A smaller matching radius would decrease the number of spurious matches, but it may also cause us to lose some real matches. To test the impact of this, we ran P3 with a higher resolution than normal: $R\sbr{g}=\sim0.1$ comoving $h^{-1}$Mpc and redshift slices having thickness of $z\sbr{g} =  0.01$. This resolution allows peaks to be potentially more than a single grid spacing away from a group center and still resolve as a match.

\tabref{mill225} shows the results of this high-resolution run of the P3 algorithm, when its results are matched to the halo catalogue using a transverse matching length of $r{mat}=0.25\,h^{-1}$Mpc instead of the usual $0.5\,h^{-1}$Mpc. The benefit of the smaller aperture size is much larger with this lower matching length, implying that the lower aperture size allows more precise determination of the positions of groups. Although the purity shown here is overall lower than with the larger matching length, the difference in purity relative to the Control is now larger.

\subsubsection{Richness of matched groups}

\begin{figure*}
  \centering
  \includegraphics[scale=0.43]{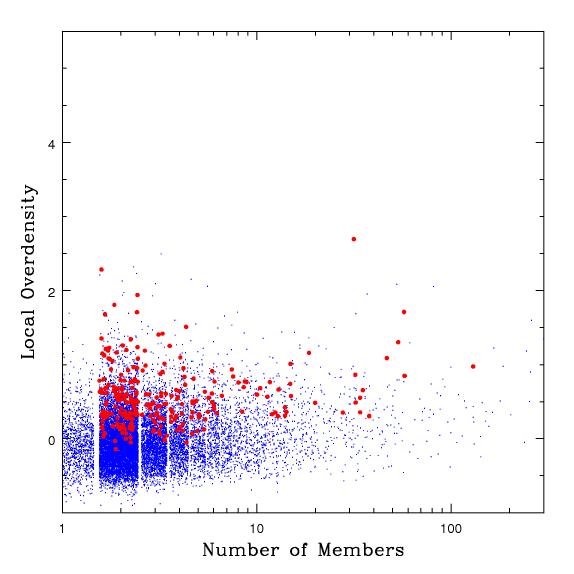}
  \hfill
  \includegraphics[scale=0.43]{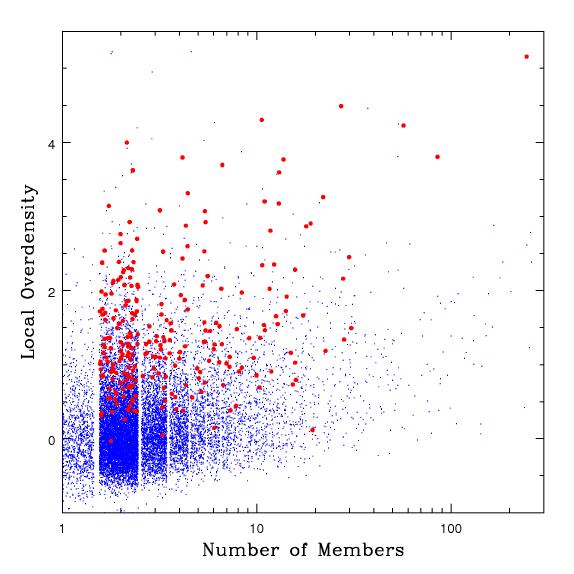}
  \hfill
  \includegraphics[scale=0.43]{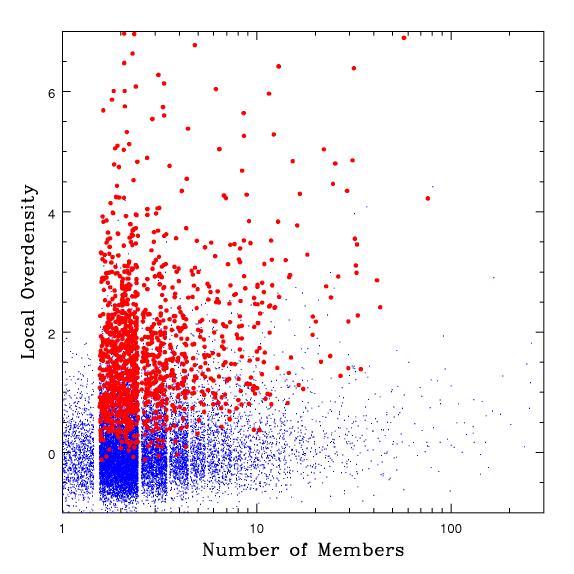}
  \hfill
  \includegraphics[scale=0.43]{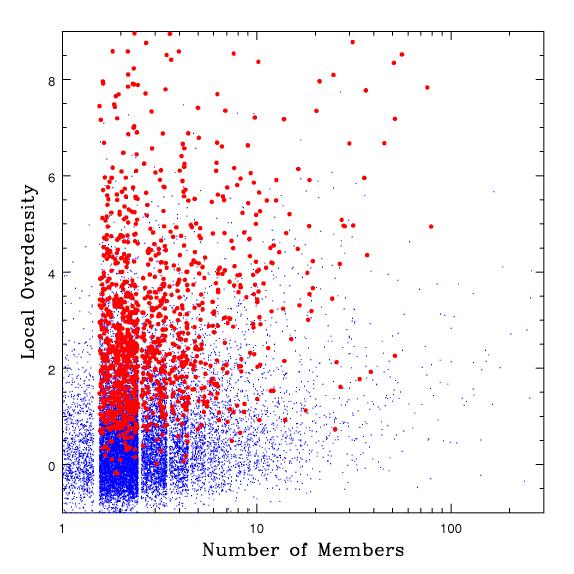}
  \hfill
  \caption[Overdensity versus number of members for data from the Millennium simulation.]{The $\delta$ calculated by our algorithm at the location of every group-containing halo (small blue dots) versus the number of members in those groups for the Millennium fields. Also shown are the $\delta$ of all estimated groups and the number of bright ({\it I} $<$ 22.5) members of the halo they match to (large red dots with error bars). The numbers of members for all data points have a random component of less than 1 included in order to aid viewing. The left column shows simulations with CFHTLSpz errors, and the right column shows simulations with COSMOS30pz errors. The top row shows simulations with $R\sbr{ap} = 0.5$ $h^{-1}$Mpc, and the bottom row shows simulations with $R\sbr{ap} = 0.25$ $h^{-1}$Mpc.}
  \label{fig:nmo}
\end{figure*}
\begin{figure}
  \centering
  \includegraphics[scale=0.38]{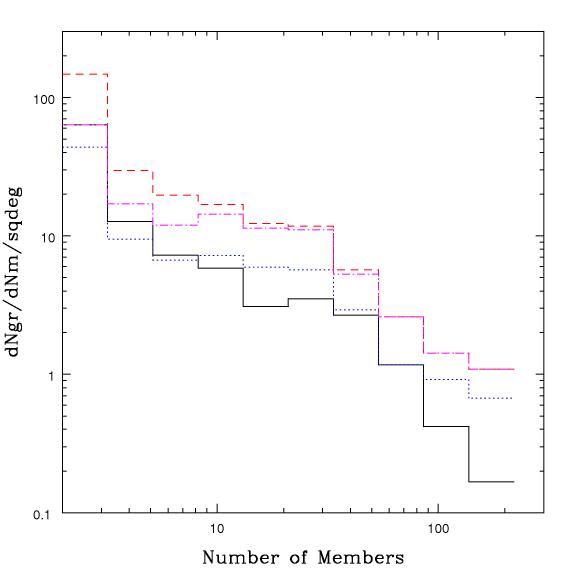}
  \hfill
  \caption[Histogram of number of groups detected with a given number of members for data from the Millennium simulation, varying aperture size and photo-z accuracy.]{The number of groups detected with a given number of members for the Millennium fields, using a minimum S/N cut of 3. Solid black: CFHTLSpz errors, $R\sbr{ap} = 0.5$ $h^{-1}$Mpc. Dotted blue: COSMOS30pz errors, $R\sbr{ap} = $ $h^{-1}$Mpc. Dashed red: CFHTLSpz errors, $R\sbr{ap} = 0.25$ $h^{-1}$Mpc. Dash-dotted magenta: COSMOS30pz errors, $R\sbr{ap} = 0.25$ $h^{-1}$Mpc.}
  \label{fig:snrhist}
\end{figure}
\begin{figure}
  \centering
  \includegraphics[scale=0.38]{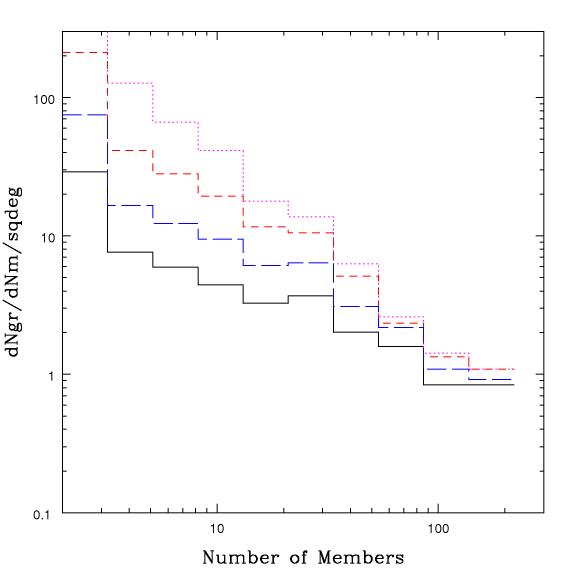}
  \hfill
  \caption[Histogram of number of groups detected with a given number of members for data from the Millennium simulation, varying minimum overdensity cut.]{The number of groups detected with a given number of members for the Millennium fields, using all peaks detected, CFHTLSpz errors, and $R\sbr{ap}$ = 0.25 $h^{-1}$Mpc, for various cuts on $\delta$. Dotted magenta: $\delta$ $>$ 2. Short dashed red: $\delta$ $>$ 3. Long dashed blue: $\delta$ $>$ 4. Solid black: $\delta$ $>$ 5.}
  \label{fig:odhist}
\end{figure}

The typical group matched by the P3 algorithm has 5-15 members, though this number depends on what signal-to-noise cut and which level of photo-z errors we used. Even though it might seem that a high signal-to-noise cut would significantly bias us toward highly populated groups, the fact that there are many more poor groups than rich groups means that many of these groups will, by chance, have a large signal-to-noise and be detected by our algorithm. This effect can easily be seen in \figref{nmo}, \figref{snrhist}, and \figref{odhist}.

There are some additional potential issues with the method we've used to assess P3's accuracy. The use of a strict matching radius between a peak location and a group centre means that we may miss some larger systems if the detected peak lies far enough from the calculated centre of the halo, even if the peak does lie within the group. This effect is lessened by using a larger matching length, but this also makes detections more prone to background contamination.

In order to estimate the richness of our detected groups, we investigated whether the local $\delta$ could be used to estimate the number of members contained in the group. \figref{nmo} shows the $\delta$ versus the number of bright ({\it I} $<$ 22.5) members in the matched halo for each photometrically-detected group that matched to at least one halo. We also calculated the $\delta$ at the location of every halo for further data.

Although there is a weak correlation between the number of members in a group and its $\delta$, using this to estimate the number of members is problematic. This is primarily due to the fact that there are many more poor groups than rich groups, and due to the large photometric uncertainties, many of these poor groups will have their $\delta$ scattered to high values. For instance, a group with $\delta = 4$ (measured with  $R\sbr{ap}=0.25\,h^{-1}$Mpc) has a roughly equal chance to have 2 members as it does to have 10. Given this, a simple mapping of $\delta$ to number of members would not likely be useful.

\begin{table}
 \centering
  \caption[Summary of group-finding accuracy for data from the Millennium simulation, testing whether removing rich groups from the catalogue affects our accuracy.]{Summary of the purity and completeness of the P3 algorithm using $R\sbr{ap}$ = 0.5 $h^{-1}$Mpc, CFHTLSpz errors (comparable to \tabref{mill}, top row, left column), and including all galaxies with {\it I} $<$ 22.5 that do not lie within a rich ($N_{m}>20$) group, matched to the halo catalogue of poor ($N_{m}\leq20$) groups, for various minimum S/N limits and a Control. The purity shows no statistically significant decrease relative to the catalogue used for \tabref{mill}, showing that rich groups are not significantly biasing our purity and completeness upwards. The combined galaxy catalogues cover a total of 12 deg.$^2$ of simulated sky. Columns are as \tabref{mill}}
  \resizebox{\columnwidth}{!}{%
  \begin{tabular}{@{}lrrrrrrrrrr@{}}
  \hline
Cut & $R\sbr{ap}$ & Nhit & Npeak & P & C & $\left<N\sbr{m}\right>$ & $\left<m\right>$ & $\left<N\sbr{mat}\right>$ & $\left<N\sbr{field}\right>$ \\
\hline
Control & 0.5 & 1705 & 2988 & 0.57 & N/A & 3.87 & 1.37 & 1.82 & 6.64 \\
All Peaks & 0.5 & 2767 & 3756 & 0.74 & 0.19 & 5.38 & 2.28 & 2.08 & 6.87 \\
S/N $>$ 2 & 0.5 & 2062 & 2669 & 0.77 & 0.14 & 6.06 & 2.37 & 2.18 & 7.24 \\
S/N $>$ 3 & 0.5 & 1299 & 1591 & 0.82 & 0.09 & 6.87 & 2.38 & 2.35 & 7.63 \\
S/N $>$ 4 & 0.5 & 585 & 690 & 0.85 & 0.04 & 8.31 & 2.37 & 2.62 & 8.21 \\
\hline
\end{tabular}}
 \label{tab:millp}
\end{table}

One potential concern was that the above-random match rate of our photometric galaxy catalogue to the halo catalogue might have been due primarily to a very high match rate among richer groups averaged with a lower match rate to poorer groups. If this were the case, then our method could in actuality be little better than random for identifying poor groups. To test this, we took our simulated galaxy catalogues and removed all galaxies within them that were found to be a member of a rich group (which we considered any group having more than 20 members to be). We then ran our algorithm on this pruned catalogue and assessed its accuracy through the same method as before. The results of this test are summarized in \tabref{millp}, which shows that there is in fact very little effect on the accuracy of our algorithm when the richer groups are removed from consideration. The purity in fact rises slightly with this test, which may be due to P3 detecting the rich groups at a point removed from their centres, as these groups are possibly large enough that the distance between the peak S/N and the group center may exceed the matching length.
 
\subsubsection{Background contamination}

The best-match groups we compared our peaks to in \tabref{mill} and \tabref{mill225} may not be the only causes of the high detected S/N. It is possible that other nearby groups and field galaxies are also contributing to the S/N at these peaks. The $\left<N\sbr{mat}\right>$ and $\left<N\sbr{field}\right>$ columns in these tables give the average total number of groups within the matching distance and the average number of field galaxies within the matching distance respectively. It can be seen from this that with the large aperture size and matching length, most of our peaks actually match to 2 or more groups, with $\left<N\sbr{mem}\right>= 7.2$ and approximately 8 field galaxies also within the matching distance. Thus typically, the interloper fraction is approximately $69\%$. Both of the number of matched groups and the field contamination decrease when lowering the matching length, but so does the measured purity. For the smaller aperture size, the lower matching length is likely a more reasonable measurement, as galaxies outside this smaller length won't be contributing to the measured S/N.

\subsection{Optimization of the algorithm}

\subsubsection{Aperture size ($R\sbr{ap})$}

\begin{figure*}
  \centering
  \includegraphics[scale=0.4]{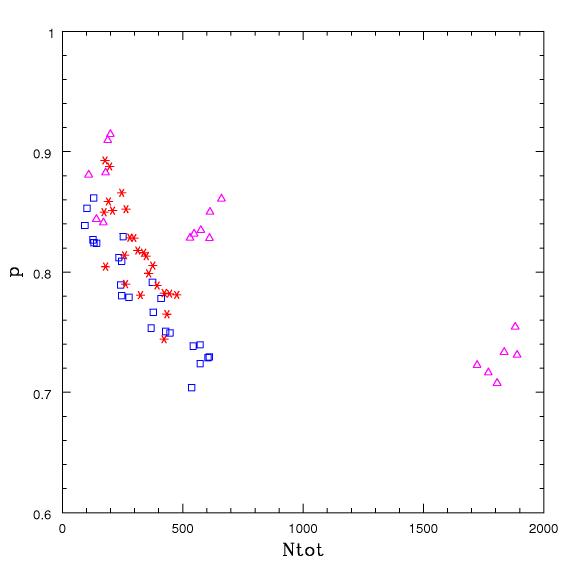}
  \centering
  \includegraphics[scale=0.4]{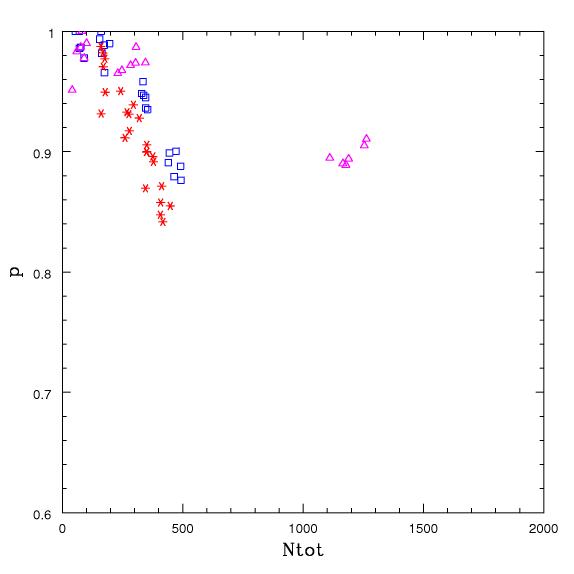}
  \caption[Purity versus number of groups detected for various settings and fields.]{Purity as a function of the total number of peaks detected. Left plot shows simulations with CFHTLSpz errors, right plot shows simulations with COSMOS30pz errors. Blue squares: $R\sbr{ap}$ = 0.5 and {\it I} $<$ 22.5. Magenta triangles: $R\sbr{ap}$ = 0.25 and {\it I} $<$ 22.5. Red stars: $R\sbr{ap}$ = 0.5 and {\it I} $<$ 24 (discussed in \secref{faintdat}). Each field from the Millennium simulation is represented by three points, for group catalogues limited by S/N $>$ 2, S/N $>$ 3, and S/N $>$ 4.}
  \label{fig:nmp}
\end{figure*}

Although decreasing the aperture size typically resulted in decreasing the purity of our catalogue at a given S/N cut, it also greatly increased the number of peaks detected. The important measurement is whether the decreased aperture size results in increased purity when the same number of groups are detected, or similarly, whether the decreased aperture size results in more groups detected at the same purity level. \figref{nmp} shows a graphical representation of how the purity relates to the number of groups detected for both aperture sizes, along with the results of changing to a fainter magnitude limit (discussed below in \secref{faintdat}). From this plot, it is clear that the smaller aperture size is beneficial, though it may require a larger S/N cut to attain sufficient purity.

\subsubsection{Magnitude limit}
\label{sec:faintdat}

Although galaxies with {\it I} $>$ 22.5 have less accurate redshifts, they should still provide some redshift information that could be useful for identifying groups. In theory, poor-precision redshift data can be used to increase the accuracy of group identifications.

\begin{table*}
 \centering
  \caption[Summary of group-finding accuracy for data from the Millennium simulation, using fainter galaxy data.]{Summary of the purity and completeness of the P3 algorithm applied to the Millennium fields with $R\sbr{ap}$ = 0.5 (comparable to \tabref{mill}, top row), and including all galaxies with {\it I} $<$ 24 when matched to the halo catalogue of groups, containing 31668 total groups, for various minimum S/N limits and a Control. The combined galaxy catalogues cover a total of 12 deg.$^2$ of simulated sky. Columns are as \tabref{mill}.}
  \resizebox{\textwidth}{!}{%
  \begin{tabular}{@{}lrrrrrrrrrrrrrrrrr@{}}
  \hline
   & & \multicolumn{8}{c}{CFHTLSpz errors} & \multicolumn{8}{c}{COSMOS30pz errors} \\
  \hline
  Cut & $R\sbr{ap}$ & Nhit & Npeak & P & C & $\left<N\sbr{m}\right>$ & $\left<m\right>$ & $\left<N\sbr{mat}\right>$ & $\left<N\sbr{field}\right>$ & Nhit & Npeak & P & C & $\left<N\sbr{m}\right>$ & $\left<m\right>$ & $\left<N\sbr{mat}\right>$ & $\left<N\sbr{field}\right>$ \\
  \hline 
Control & 0.5 & 1705 & 2988 & 0.57 & N/A & 3.87 & 1.37 & 1.82 & 6.64 & 1705 & 2988 & 0.57 & N/A & 3.87 & 1.37 & 1.82 & 6.64 \\
All Peaks & 0.5 & 2006 & 2592 & 0.77 & 0.15 & 7.16 & 3.28 & 2.16 & 6.32 & 2125 & 2470 & 0.86 & 0.19 & 7.43 & 3.46 & 2.36 & 6.61 \\
S/N $>$ 2 & 0.5 & 1655 & 2055 & 0.81 & 0.13 & 7.86 & 3.52 & 2.24 & 6.54 & 1878 & 2088 & 0.90 & 0.17 & 8.00 & 3.74 & 2.44 & 6.85 \\
S/N $>$ 3 & 0.5 & 1333 & 1607 & 0.83 & 0.10 & 8.80 & 3.74 & 2.30 & 6.72 & 1506 & 1619 & 0.93 & 0.14 & 9.07 & 4.19 & 2.58 & 7.20 \\
S/N $>$ 4 & 0.5 & 964 & 1124 & 0.86 & 0.08 & 10.52 & 4.07 & 2.44 & 7.19 & 979 & 1013 & 0.97 & 0.09 & 11.78 & 5.10 & 2.81 & 7.75 \\
  \hline
\end{tabular}}
 \label{tab:millf}
\end{table*}

We ran the P3 algorithm on our simulated galaxy catalogue, using all galaxies with {\it I} $<$ 24, assigning errors to galaxies with 22.5 $<$ {\it I} $<$ 24 of twice as much as for galaxies with {\it I} $<$ 22.5, resulting in 0.10 for the CFHTLSpz errors and 0.04 for the COSMOS30pz errors. In the end, using faint galaxy photometric redshifts showed a small decrease in the purity of group-finding for CFHTLSpz errors, but a small increase for COSMOS30pz errors, as can be seen in a comparison of \tabref{millf} and \tabref{mill}. As a result, it will not be worth the extra computation time to include galaxies with photo-z errors $>0.05$ in runs of P3 on real data.

\begin{figure*}
  \centering
  \centering
  \includegraphics[scale=0.5]{millay382.jpeg}
  \centering
  \includegraphics[scale=0.5]{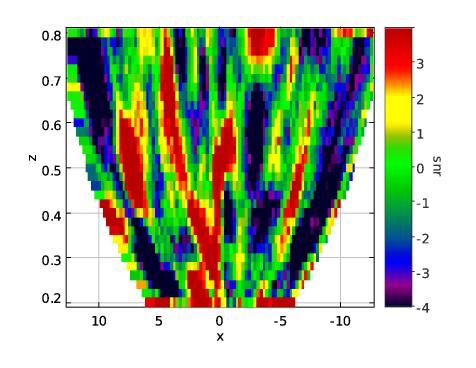}
  \caption[S/N plots in x and z for data from the Millennium simulation, showing the effect of also using fainter galaxy data.]{S/N maps generated by running P3 on the simulated catalogue with CFHTLSpz errors and $R\sbr{ap}=0.5\,h^{-1}$Mpc. Left plot was generated from a run which included all galaxies with {\it I} $<$ 22.5, and the right plot was generated from a run which included all galaxies with {\it I} $<$ 24. Galaxies with 22.5 $<$ {\it I} $<$ 24 were assigned errors of twice as much as the brighter galaxies. It can be seen here that including fainter galaxies tends to increase the contrast of the distribution by increasing the S/N in apparent structures, but there is no significant change to the shape of it.}
  \label{fig:millfsnry}
\end{figure*}

As the locations of fainter galaxies are highly correlated with the locations of bright galaxies, the effect of using them in the P3 algorithm is to increase the contrast of the existing S/N distribution, albeit at a lower resolution in the redshift dimension, as can be seen in \figref{millfsnry}. The lower redshift resolution also manifests in a greater error in determination of the peak redshift, which may result in more peaks being scattered away from the group they represent by more than $z\sbr{mat}$. It is also possible that multiple groups along the line of sight may become blended into a single structure, with the peak of this structure lying between the two groups. When the peak catalogue is compared to the actual locations of the groups, it is possible that it is within our threshold matching distance of neither.

The result of our optimization analysis is that the $0.25\,h^{-1}$Mpc aperture size performs better, and a S/N cut of 3 provides a good balance of purity and completeness without overly biasing us toward rich groups. A resolution of $\le0.1\,h^{-1}$Mpc is preferred with this aperture size, although it is more computationally expensive. The magnitude limit for a real galaxy catalogue will depend on the distribution of $\sigma_z$ vs. $z$. A magnitude which limits $\sigma_z$ to a maximum of $0.05$ will likely provide the best results, as it was when the simulated errors for fainter galaxies went above this level that the purity began to decrease.

\subsection{Comparison to mock spectroscopic ``Friends-of-friends'' groups}

\begin{table*}
 \centering
  \caption[Summary of group-finding accuracy for FoF catalogue from the Millennium simulation.]{Summary of the purity and completeness of the P3 algorithm when its results are matched to a group catalogue obtained through a Friends-of-Friends algorithm applied to galaxies from the Millennium simulation using their given redshifts, containing 39101 total FoF groups, for various minimum S/N limits and a Control. The combined galaxy catalogues cover a total of 12 deg.$^2$ of simulated sky. Columns are as \tabref{mill}.}
  \resizebox{\textwidth}{!}{
  \begin{tabular}{ l r | r r r r r r r r | r r r r r r r r }
  \hline
   & & \multicolumn{8}{c|}{CFHTLSpz errors} & \multicolumn{8}{c}{COSMOS30pz errors} \\
  \hline
  Cut & $R\sbr{ap}$ & Nhit & Npeak & P & C & $\left<N\sbr{m}\right>$ & $\left<m\right>$ & $\left<N\sbr{mat}\right>$ & $\left<N\sbr{field}\right>$ & Nhit & Npeak & P & C & $\left<N\sbr{m}\right>$ & $\left<m\right>$ & $\left<N\sbr{mat}\right>$ & $\left<N\sbr{field}\right>$ \\
  \hline
Control & 0.5 & 1960 & 2988 & 0.66 & N/A & 3.68 & 0.76 & 2.19 & 4.02 & 1960 & 2988 & 0.66 & N/A & 3.68 & 0.76 & 2.19 & 4.02 \\
All Peaks & 0.5 & 2643 & 3439 & 0.77 & 0.17 & 4.31 & 1.30 & 2.65 & 4.85 & 2549 & 2805 & 0.91 & 0.21 & 4.42 & 1.58 & 3.06 & 5.76 \\
S/N $>$ 2 & 0.5 & 1920 & 2406 & 0.80 & 0.13 & 4.77 & 1.40 & 2.84 & 4.98 & 1951 & 2048 & 0.95 & 0.17 & 4.93 & 1.80 & 3.40 & 6.07 \\
S/N $>$ 3 & 0.5 & 1238 & 1496 & 0.83 & 0.09 & 5.39 & 1.55 & 3.07 & 5.17 & 1017 & 1025 & 0.99 & 0.10 & 6.53 & 2.31 & 4.17 & 6.47 \\
S/N $>$ 4 & 0.5 & 632 & 725 & 0.87 & 0.05 & 6.86 & 1.78 & 3.32 & 5.36 & 413 & 415 & 1.00 & 0.05 & 9.35 & 3.64 & 4.80 & 6.71 \\
\hline
Control & 0.25 & 1960 & 2988 & 0.66 & N/A & 3.68 & 0.76 & 2.19 & 4.02 & 1960 & 2988 & 0.66 & N/A & 3.68 & 0.76 & 2.19 & 4.02 \\
All Peaks & 0.25 & 13791 & 20727 & 0.67 & 0.65 & 3.68 & 0.98 & 2.15 & 4.05 & 13256 & 18724 & 0.71 & 0.67 & 3.72 & 0.99 & 2.20 & 4.12 \\
S/N $>$ 2 & 0.25 & 8637 & 10904 & 0.79 & 0.44 & 4.28 & 1.12 & 2.44 & 4.40 & 6662 & 7160 & 0.93 & 0.39 & 4.84 & 1.36 & 2.75 & 4.75 \\
S/N $>$ 3 & 0.25 & 3152 & 3537 & 0.89 & 0.18 & 6.23 & 1.35 & 3.00 & 4.83 & 1688 & 1715 & 0.98 & 0.13 & 8.76 & 2.25 & 3.72 & 5.39 \\
S/N $>$ 4 & 0.25 & 906 & 986 & 0.92 & 0.06 & 10.18 & 1.68 & 3.54 & 5.22 & 444 & 448 & 0.99 & 0.04 & 15.81 & 3.02 & 4.04 & 5.79 \\
  \hline
  \end{tabular}}
 \label{tab:millfof}
\end{table*}

In the following section we will compare the P3 group catalogue to spectroscopically-identified groups from zCOSMOS. When working with real-world data, peculiar velocities of galaxies in groups cause small redshift errors, which any group-finding method must adapt for. The result of this is that spectroscopic identification of groups suffers from its own imperfections in purity and completeness. A comparison of the P3 group catalogue to a spectroscopic catalogue will let us assess how much imperfections in real galaxy catalogues will affect group-finding.

To make this assessment, we first need to determine how much the purity and completeness drops when we switch from a comparison with the halo positions to a comparison with a catalogue that better simulates what we might obtain with real galaxy catalogues. For this purpose, we ran a friends-of-friends algorithm using the redshifts of galaxies in the Millennium simulation. The generated catalogue contained a total of 39101 groups with at least two bright members, with approximately 3300 groups per deg.$^2$ This are approximately $25\%$ more groups than the halo catalogue, and the effect of this can be seen in the increased 'purity' of the Control catalogue. The increase in the Control purity is less than $25\%$, which is likely due to many of the new FoF groups lying close together, as would happen if some real groups were detected as multiple FoF groups. A Control peak which lies within the matching threshold of both of these groups is only counted as a single match, so $25\%$ more groups, some of which lie near other groups, will result in a purity increase of less than $25\%$. Interestingly, the average number of field galaxies detected near each peak dropped, which may be due to the FoF algorithm matching some field galaxies into spurious pairs.

\tabref{millfof} shows the results of a comparison with this catalogue. The Control 'purity' for this group catalogue has increased by $\sim9\%$ relative to the halo catalogue, the measured purity for peak catalogs has increased by $\sim3-7\%$, the average number of galaxies in each matched group has decreased by $\sim20-40\%$, and the average number of groups matched has increased by $\sim20-35\%$. This result is most likely caused by the FoF algorithm fragmenting real groups into multiple FoF groups. The result of this would be a higher density of groups with fewer members per group, consistent with the observed results. The lower number of field galaxies matched on average, however, is not explained by groups being fragmented. This is possibly a result of field galaxies being grouped into spurious pairs by the FoF algorithm. Both of these hypotheses are consistent with the observation, as can be seen in \figref{gnumcomp} below, that the FoF catalogue contains more poor groups than the halo catalogue and fewer rich groups.

\subsection{zCOSMOS Galaxy Catalogues}
\label{sec:realdat}

In addition to simulated galaxy catalogues, we tested the P3 algorithm on galaxy catalogues from the COSMOS/CFHTLS D2 field. This field has spectroscopic redshifts for a large number of the galaxies, in addition to photometric redshifts based on {\it u*griz} photometry from \citet{IlbArnMcC06} with errors of $\sigma_z\sim0.05$ for {\it i'} $<$ 22.5, and also much smaller errors from the COSMOS 30 band photometry \citep{IlbCapSal09} ($\sigma_z\sim0.02$ for {\it i'} $<$ 22.5). This allows us to better see what purity we can expect from when we run the P3 algorithm on the CFHTLS-Wide survey, and how the accuracy might improve in surveys with better photometric redshift accuracy.

We have used the spectroscopic group catalogue by \citet{KnoLilIov09} to assess the accuracy of our method. The catalogue contains 604 groups in redshifts between $z\sbr{lo}=0.2$ and $z\sbr{hi}=0.8$ over 1.7 deg.$^2 $, which is only $\sim13\%$ as many groups as were detected in the same area with the simulated galaxy catalogues. The discrepancy is not accounted for by the completeness of this catalogue, claimed to be $85\%$ by \citeauthor{KnoLilIov09} for sampled galaxies, with a $70\%$ sampling rate. \citep{LilLeFRen07} The discrepancy can be seen illustrated in \figref{gnumcomp}, which shows \citeauthor{KnoLilIov09}'s group counts for varying richness, corrected for completeness, sampling rate, and the differing galaxy density of their D2 field from the Millenium fields, compared to our group catalogues.

The discrepancy is better explained by the lower completeness of \citeauthor{KnoLilIov09}'s group-finding method for poor groups. As can be seen from Fig. 2 from \citet{KnoLilIov09}, for groups with 10 or few members, which have a typical mass of $10^{13}\,h^{-1}M_{\odot}$, the completeness is less than $40\%$. Additionally, \citeauthor{KnoLilIov09} noted that the D2 field appears to have an unusually low density of rich groups, which explains the cut-off seen in \figref{gnumcomp}.

In comparing the photometric redshifts to spectroscopic redshifts, we found that there appeared to be a small, but significant, offset between the photometric redshifts provided by \citet{IlbArnMcC06} and the spectroscopic redshifts. To correct for this, we performed a sky match on galaxies present and both of the catalogues and fit a linear correction function to their redshifts, of the form $z_{real} = 0.957z_{phot}+0.00843$. This correction allowed us to properly match our detected groups to those from \citeauthor{KnoLilIov09}. Although this correction had a significant effect on the apparent quality of our group-matching, it is unlikely to have any significant effect on lensing measurements.

\begin{figure*}
  \centering
  \includegraphics[scale=0.5]{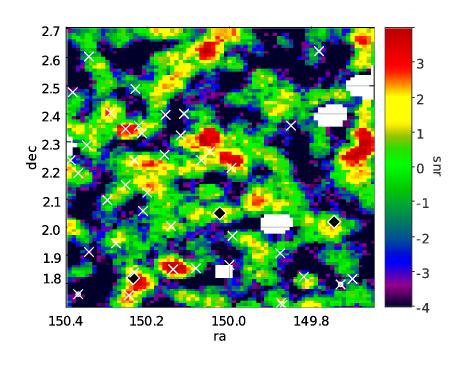}
  \hfill
  \includegraphics[scale=0.5]{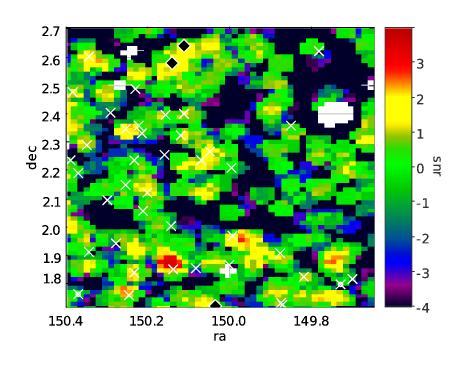}
  \hfill
  \includegraphics[scale=0.5]{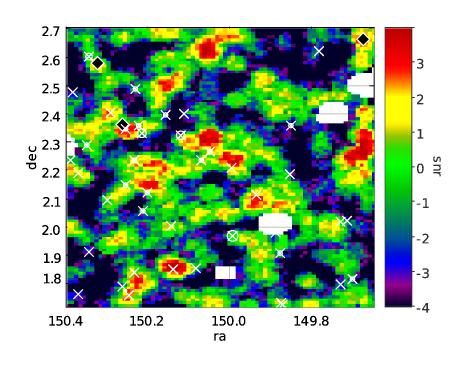}
  \hfill
  \includegraphics[scale=0.5]{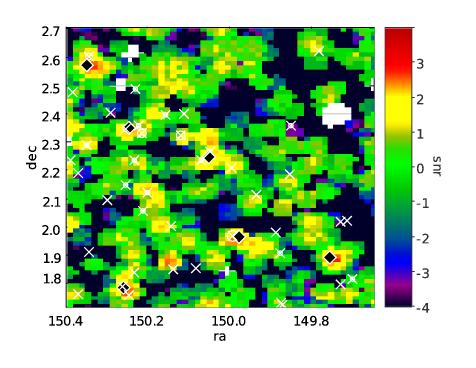}
  \hfill
  \includegraphics[scale=0.5]{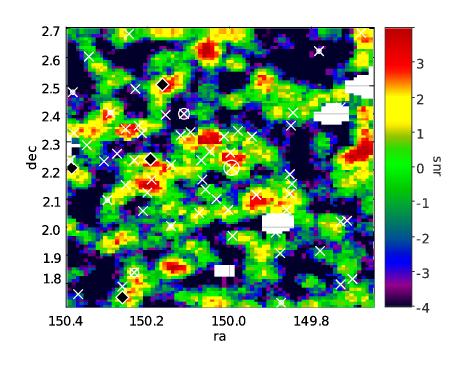}
  \hfill
  \includegraphics[scale=0.5]{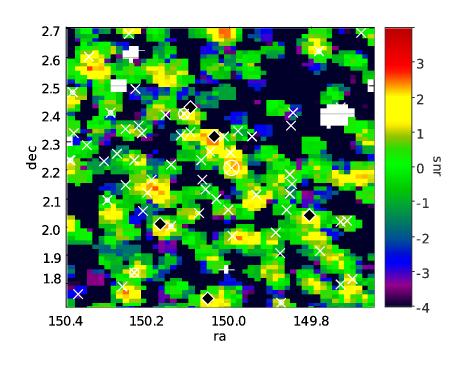}
  \hfill
  \caption[S/N plots in R.A. and Dec. for data from the COSMOS D2 field.]{The calculated S/N for the $\delta$ of galaxies on a grid of points in R.A. and Dec., sliced at different values of the redshift, for the D2 field, using $R\sbr{ap}$ = 0.5 $h^{-1}$Mpc. S/N is indicated by the colour. Locations of real groups detected through \citeauthor{KnoLilIov09}'s friends-of-friends algorithm are indicated by white circles, with their sizes indicating the sizes of the groups. White crosses indicate the location of a circle in a nearby layer, within our threshold redshift for being considered a match. Detected peaks with a S/N of more than 2 are indicated by the black diamonds. Peaks are detected in three dimensions, so what appear to be peaks in the individual plots may actually be detected on another slice. Additionally, peaks have a threshold radius within which they must be the highest point to count as a peak, so some peaks may not be detected if they are sufficiently close to another peak. Left column shows data derived from the \citet{IlbArnMcC06} photo-zs, with errors similar to the CFHTLSpz errors used previously, and the right column shows data derived from the COSMOS-30 photo-zs, with errors similar to the COSMOS30pz erros used previously. Redshift slices, from top to bottom: 0.58, 0.60, 0.62. The COSMOS plot shows the interesting effect that many galaxies are individually resolvable, as their redshift errors are smaller than the width of the slices. These galaxies appear as solid circles in the plot.}
  \label{fig:d2snrz}
\end{figure*}
\begin{figure*}
  \centering
  \includegraphics[scale=0.5]{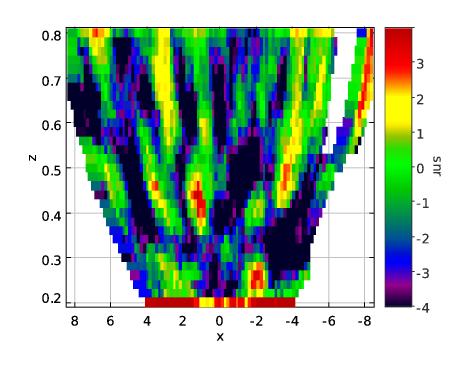}
  \centering
  \includegraphics[scale=0.5]{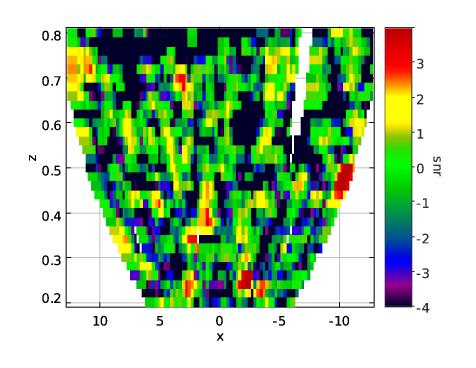}
  \caption[S/N plots in x and z for data from the COSMOS D2.]{An alternate view of the plots from \figref{d2snrz}, using a slice at constant Dec., showing how far groups extend in the redshift dimension. The left plot was constructed with \citet{IlbArnMcC06}'s photo-zs and $R\sbr{ap} = 0.5$ $h^{-1}$Mpc; the right was constructed with the COSMOS-30 photo-zs and $R\sbr{ap} = 0.5$ $h^{-1}$Mpc. Note the differing scale in the comparison, as the COSMOS-30 catalogue covers a somewhat larger area. Here, we can clearly see in the COSMOS-30 plot the contraction of groups in the redshift dimension.}
  \label{fig:d2snry}
\end{figure*}

\figref{d2snrz} and \figref{d2snry} show a graphical representation of the S/N in selected redshift slices for the D2 field. Although the catalogue of FoF-identified groups is significantly sparser, most P3 groups do correspond to a FoF-identified group.

\begin{table*}
 \centering
  \caption[Summary of group-finding accuracy for data from the COSMOS D2 field.]{Summary of the purity and completeness of the P3 algorithm applied to the galaxies in the D2 field. The combined galaxy catalogues cover approximately 1 deg.$^2$ of sky. Columns are as \tabref{mill}. Masses were calculated through an abundance matching technique by \citeauthor{KnoLilIov09}.}
  \resizebox{\textwidth}{!}{%
  \begin{tabular}{@{}lrrrrrrrrrrrrrrr@{}}
  \hline
   & & \multicolumn{7}{c}{CFHTLS photometry} & \multicolumn{7}{c}{COSMOS-30 photometry} \\
  \hline
  Cut & $R\sbr{ap}$ & Nhit & Npeak & P & C & $\left<N\sbr{m}\right>$ & $\left<m\right>$ & $\left<N\sbr{mat}\right>$ & Nhit & Npeak & P & C & $\left<N\sbr{m}\right>$ & $\left<m\right>$ & $\left<N\sbr{mat}\right>$ \\
  \hline
Control & 0.5 & 55 & 200 & 0.28 & N/A & 3.42 & 5.71 & 1.44 & 55 & 200 & 0.28 & N/A & 3.42 & 5.71 & 1.44 \\
All Peaks & 0.5 & 53 & 100 & 0.53 & 0.20 & 3.36 & 7.26 & 1.62 & 94 & 157 & 0.60 & 0.38 & 3.30 & 6.82 & 1.63 \\
S/N $>$ 2 & 0.5 & 46 & 77 & 0.60 & 0.17 & 3.41 & 7.60 & 1.67 & 75 & 98 & 0.77 & 0.31 & 3.55 & 7.32 & 1.75 \\
S/N $>$ 3 & 0.5 & 35 & 50 & 0.70 & 0.13 & 3.43 & 6.76 & 1.77 & 35 & 40 & 0.88 & 0.18 & 4.11 & 7.78 & 2.11 \\
S/N $>$ 4 & 0.5 & 16 & 21 & 0.76 & 0.06 & 3.81 & 6.39 & 2.06 & 13 & 13 & 1.00 & 0.08 & 3.38 & 6.09 & 2.46 \\
\hline
Control & 0.25 & 55 & 200 & 0.28 & N/A & 3.42 & 5.71 & 1.44 & 55 & 200 & 0.28 & N/A & 3.42 & 5.71 & 1.44 \\
All Peaks & 0.25 & 230 & 605 & 0.38 & 0.61 & 3.26 & 6.13 & 1.28 & 322 & 1148 & 0.28 & 0.79 & 3.15 & 6.01 & 1.28 \\
S/N $>$ 2 & 0.25 & 156 & 321 & 0.49 & 0.41 & 3.65 & 7.53 & 1.30 & 112 & 147 & 0.76 & 0.34 & 3.82 & 9.22 & 1.50 \\
S/N $>$ 3 & 0.25 & 59 & 91 & 0.65 & 0.18 & 3.75 & 6.91 & 1.41 & 19 & 20 & 0.95 & 0.09 & 4.79 & 12.93 & 1.74 \\
S/N $>$ 4 & 0.25 & 9 & 12 & 0.75 & 0.03 & 4.67 & 7.48 & 1.33 & 0 & 1 & 0.00 & 0.01 & N/A & N/A & N/A \\
  \hline
\end{tabular}}
 \label{tab:d2}
\end{table*}

\begin{figure}
  \centering
  \includegraphics[scale = 0.4]{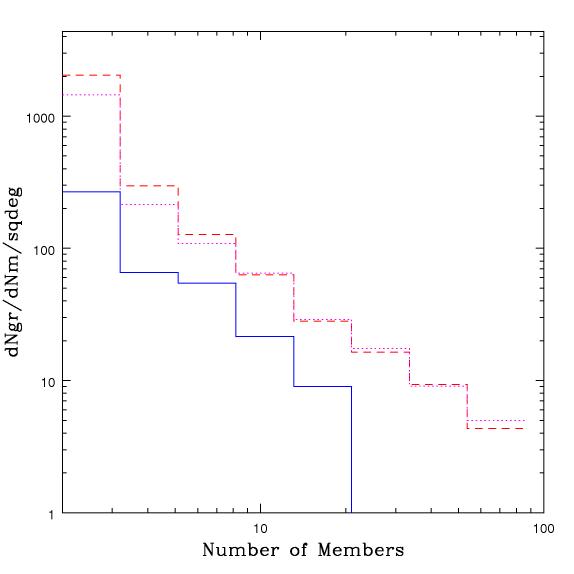}
  \caption[Completeness comparison for Millennium and COSMOS groups detected.]{A comparison the number of groups of varying numbers of members identified in the magnitude range {\it i'} $<$ 22.5 for both the Millennium simulation generated through halo catalogues (magenta dotted) and through a friends-of-friends algorithm (red dashed), and for the zCOSMOS FoF-groups of \citet{KnoLilIov09} (blue solid). We can assess completeness by measuring how far the zCOSMOS plot lies to the left of the Millennium plot. This gives us an estimate of 20-40\% completeness among the groups detected.}
  \label{fig:gnumcomp}
\end{figure}

As can be seen in \tabref{d2}, P3's purity using real photometric redshifts and spectroscopically-identified groups is somewhat worse than its purity from using simulated galaxy catalogues with CFHTLSpz errors, as would be expected from the inaccuracies inherent in real-world data. Additionally, the group catalogue suffers from incompleteness due to the inaccuracy of the group-finding algorithm and the incomplete sampling of galaxies in the field. As mentioned previously, the completeness of this group catalogue for groups of mass less than $10^{14}\,M_{\odot}$ is $\sim40\%$. These completeness effects are also evident in the smaller average number of members in our detected groups, compared to the averages for the simulated catalogues. They are also evident in the fact that the Control catalogue shows lower 'purity' for this field than for the Millennium fields. \figref{gnumcomp} compares the group counts for various memberships of the D2 and Millennium fields, directly illustrating the incompleteness of the D2 galaxy and group catalogues.

\begin{figure*}
  \centering
  \includegraphics[scale=0.38]{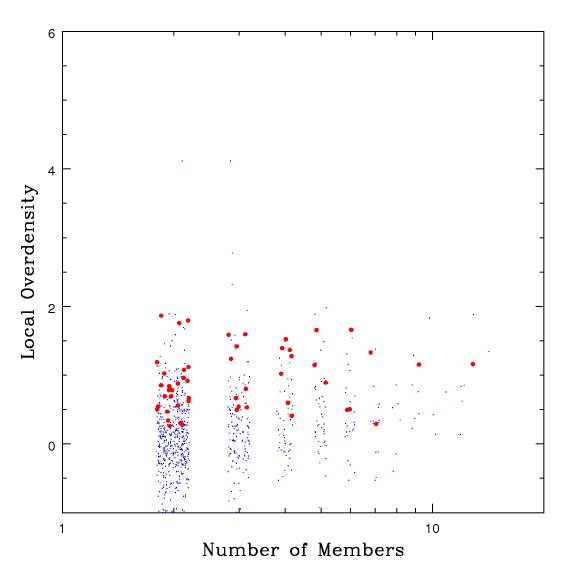}
  \hfill
  \includegraphics[scale=0.38]{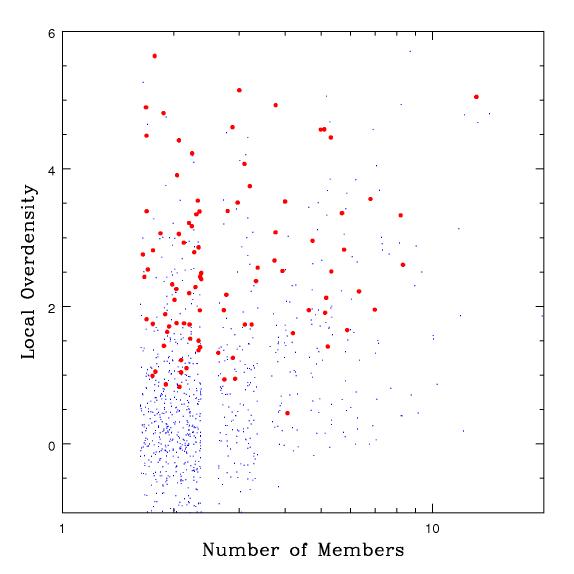}
  \caption[Overdensity versus number of members for data from the COSMOS D2 field.]{The overdensity $\delta$ calculated by the P3 algorithm at the location of every FoF-identified group (small dots) versus the number of members in those groups for the D2 field. Also shown are the $\delta$ of all estimated groups and the number of bright ({\it i'} $<$ 22.5) members of the FoF-identified group they match to (large dots with error bars). The numbers of members for all data points have a random component of less than 1 included in order to aid viewing. Left plot shows simulations with photometry from \citet{IlbArnMcC06}, right plot shows simulations with photometry from the COSMOS-30 survey.}
  \label{fig:d2nmo}
\end{figure*}

We also attempted to assess how accurately we will be able to determine the size of a group from its local $\delta$ using these catalogues, illustrated in \figref{d2nmo}. Although the group catalogue is significantly sparser than the catalogues extracted from the Millennium simulation, a positive correlation between $\delta$ and the number of members in a group can still be seen. However, the trend is not significant enough to allow us to make future estimates of the richness of groups from their $\delta$.

\section{Preliminary Application to the CFHTLS-Wide}

The CFHTLS-Wide survey is a 170 deg.$^2$ survey over four patches of sky, taken by the Canada-France-Hawaii Telescope. Photometric
redshifts for the CFHTLS-Wide were prepared with the methods described in \citet{ErbHilLer09} and \citet{HilPieErb09} in the framework of the CFHTLenS collaboration. The catalogues will be described in detail in a forthcoming paper (Hildebrandt et al. in preparation). The photo-z's are based on the publicly available BPZ code  \citep{Ben00}, yielding an accuracy of $\sigma_z \sim 0.03$ for {\it i'} $<$ 22.5. Lensing-quality shear measurements for galaxies within the survey are currently in preparation.

We used the following parameters for our preliminary run of P3 on the currently-available data from the Wide survey:

\begin{itemize}
\item $R\sbr{ap}=0.25$ Mpc. There was a significant benefit to the smaller aperture size when P3 was applied to simulated galaxy catalogues. This wasn't the case with real catalogues, but that may be due to a low sampling rate.
\item {\it i'} $< 22.5$. Tests of P3 which included galaxies in the range $22.5 <$ {\it I} $< 24$ showed a small decrease in purity for comparable numbers of groups detected.
\item S/N $>$ 3. This limit typically provided the best balance of purity and number of groups detected.
\end{itemize}

\begin{figure*}
  \centering
  \includegraphics[scale=0.4]{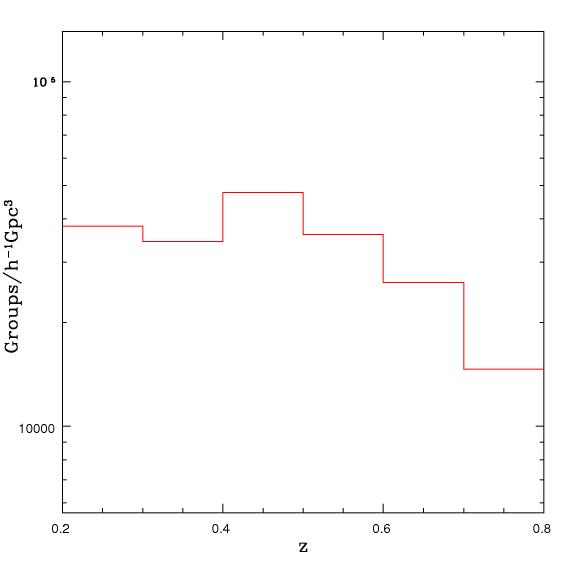}
  \hfill
  \includegraphics[scale=0.4]{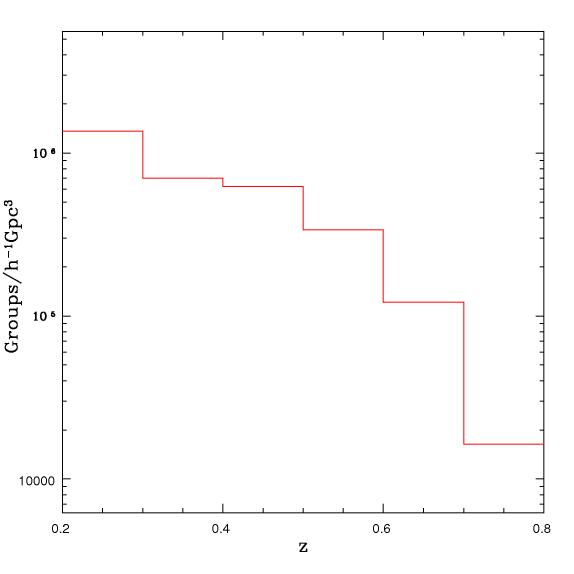}
  \caption[Histograms of the number of groups detected over redshift.]{Number of groups (peaks with S/N $>$ 3) detected over redshift. Left plot uses $R\sbr{ap} = 0.5\,h^{-1}$ Mpc, right plot uses $R\sbr{ap} = 0.25\,h^{-1}$ Mpc. Vertical axis has been scaled to give groups per $h^{-1}$ Gpc$^3$. We appear to be getting more incomplete with redshift, much more notably with the smaller aperture size. This is due to the apparent magnitude limit of the catalogue reducing the number of galaxies detected for higher-redshift groups, which decreases the chance that a given group will be detected by P3.}
  \label{fig:zhist}
\end{figure*}

With these parameters, we detected a total of 18813 groups over the 78 fields available. The fields have an average unmasked area of $\sim0.95$ deg$^2$., giving an average of 241 groups/deg$^2$. We expect approximately 80$\%$ of these detected groups to correspond to real groups. This purity is comparable to the similar work done by \citet{AdaDurBen10}, but the detected group density is much higher than their 19.2 clusters/deg$^2$.

The distribution of groups as a function of redshift is shown in \figref{zhist}. P3's group count gets more incomplete with increasing redshift, and this effect is more pronounced when using the smaller aperture size. The likely cause of this is that, at low redshifts, we can detect groups with just a few members. If one of these groups were moved to a higher redshift, fewer members would be detected, and the group itself would fall below the threshold for detection. The larger aperture size detects only richer groups in the first place, and these groups are more likely to have enough members that would still be visible were the group at a higher redshift.

\begin{table}
 \centering
  \caption[Summary of group-finding accuracy compared to \citeauthor{LuGilBal09}'s cluster catalog for the CFHTLS-Wide survey.]{Summary of the purity and completeness of the P3 algorithm applied to the available CFHTLS-Wide fields with $R\sbr{ap}$ = 0.5 Mpc to the cluster catalogue provided by \citeauthor{LuGilBal09} Our catalogue shows an above random correlation to \citeauthor{LuGilBal09}'s catalogue, with approximately 31\% purity and 69\% completeness for $R\sbr{ap} = 0.25\,h^{-1}$ Mpc and a S/N cut of 3. Columns are as \tabref{mill}, with $\left<N\sbr{m}\right>$ referring to the average number of red sequence galaxies in detected and matched groups.}
  \begin{tabular}{@{}lrrrrrrrr@{}}
  \hline
Cut & $R\sbr{ap}$ & Nhit & Npeak & P & C & $\left<N\sbr{m}\right>$ \\
\hline
Control & 0.5 & 971 & 8661 & 0.11 & N/A & 6.58 \\
All Peaks & 0.5 & 1284 & 3039 & 0.42 & 0.45 & 7.77 \\
S/N $>$ 2 & 0.5 & 1018 & 2256 & 0.45 & 0.46 & 7.72 \\
S/N $>$ 3 & 0.5 & 914 & 1701 & 0.54 & 0.41 & 7.95 \\
S/N $>$ 4 & 0.5 & 829 & 1225 & 0.68 & 0.30 & 9.01 \\
\hline
Control & 0.25 & 971 & 8661 & 0.11 & N/A & 6.58 \\
All Peaks & 0.25 & 5729 & 25301 & 0.23 & 0.82 & 6.36 \\
S/N $>$ 2 & 0.25 & 3102 & 13165 & 0.24 & 0.81 & 6.68 \\
S/N $>$ 3 & 0.25 & 2524 & 8063 & 0.31 & 0.69 & 7.11 \\
S/N $>$ 4 & 0.25 & 2014 & 4284 & 0.47 & 0.45 & 8.15 \\
\hline
\end{tabular}
 \label{tab:Lu}
\end{table}

We compared the results of P3 to \citet{LuGilBal09}'s cluster catalogue for the Wide fields, as shown in \tabref{Lu}, finding a purity of 31\%. Although this purity is low compared to our previous results, this is to be expected. As \citeauthor{LuGilBal09}'s catalogue was derived through searching for galaxies on the red sequence, it consists primarily of clusters of 10 or more large red galaxies, while our catalogue contains a large number of poorer groups that we would expect to not match to anything in \citeauthor{LuGilBal09}'s catalogue, resulting in a decreased apparent purity. The important point is that our purity is significantly above the 11\% attained with our Control catalogue. Our completeness here is 69\%, which raises the question of why, if \citeauthor{LuGilBal09}'s groups are some of the richest visible, they weren't detected by our algorithm. An inspection of S/N maps generated from the Wide catalogues showed that the groups P3 didn't pick up were typically missed for one of the following reasons:

\begin{itemize}
\item The groups were part of a structure which had a large extent in the sky, and P3 picked a different peak from \citeauthor{LuGilBal09}'s centre.
\item \citeauthor{LuGilBal09}'s group was near the edge of the field or the redshift range. P3 will not report a peak detected at the edge of the field or at the redshift limits, as it is impossible to know if this is the actual peak, or the real peak lies outside the analyzed range.
\item \citeauthor{LuGilBal09}'s group lay near a heavily-masked region, where a high S/N is less likely.
\end{itemize}

\section{Conclusion}

In this work we developed and tested a method identifying a very pure sample of galaxy groups using photometric redshifts. We predict that the method will result in a purity of $\sim 84\%$ for the quality of photometry present in the CFHTLS-Wide survey. Running our algorithm on the available fields, we detected an average of 241 groups per square degree field in the redshift range $0.2<z<0.8$, which will give a predicted 41000 groups once photometry from the entire, 170 square degree survey is available. From simulated data, we estimate that our groups have an average membership of $\sim 9$ bright ({\it I} $<$ 22.5) galaxies.

Our group-finding method shows a limited ability to estimate the size of detected groups through their local $\delta$. There is a positive correlation between the number of members in a group and its $\delta$, although the significantly larger number of smaller groups makes a direct estimate of the number of members impractical.

Our detected distribution of groups over redshift is consistent with past results \citep{MilWaeHey10}. Our group catalogue of the presently-available CFHTLS-Wide fields is consistent with \citet{LuGilBal09}'s catalogue, to the extent that would be expected given the differences in our methods.

\section*{Acknowledgments}
Based on observations obtained with MegaPrime/MegaCam, a joint project of CFHT and CEA/DAPNIA, at the Canada-France-Hawaii Telescope (CFHT) which is operated by the National Research Council (NRC) of Canada, the Institut National des Sciences de l’Univers of the Centre National de la Recherche Scientifique (CNRS) of France, and the University of Hawaii. This work is based in part on data products produced at TERAPIX and the Canadian Astronomy Data Centre as part of the Canada-France-Hawaii Telescope Legacy Survey, a collaborative project of NRC and CNRS.

This work was made possible by the facilities of the Shared Hierarchical Academic Research Computing Network (SHARCNET:www.sharcnet.ca) and Compute/Calcul Canada.

We acknowledge useful discussions with Martha Milkeraitis, and thank Simon Lilly and C. Knobel for making the zCOSMOS group catalog available to us, and thank Hendrik Hildebrandt, Thomas Erben and the CFHTLenS team for making the CFHTLS-Wide photometric redshifts available to us in advance of publication, acknowledging the use of CFI funded equipment under project grant \#10052. MJH acknowledges the hospitality of the Institut d’Astrophysique de Paris and the support of the IAP-UPMC visiting programme, the French ANR (Otarie), and the Canadian NSERC.

\bibliographystyle{mn2e}
\bibliography{mjh}

\bsp

\label{lastpage}

\end{document}